\documentclass[11pt]{article}

\usepackage{revsymb,bm,bbm,url}
\usepackage{natbib}
\usepackage[cp1250]{inputenc}   
\usepackage[T1]{fontenc}
\usepackage{amssymb,amsmath,amscd,amsfonts,dsfont}
\usepackage[dvips]{graphicx}
\usepackage[FIGTOPCAP,raggedright,nooneline,bf,footnotesize]{subfigure}
\usepackage[usenames]{color}
\usepackage{textcomp}

\newcommand{\ket}[1]{\left | #1 \right \rangle}
\newcommand{\bra}[1]{\left \langle #1 \right |}

\renewcommand{\eqref}[1]{Eq.~(\ref{#1})}

\newcommand{\figref}[1]{Fig.~\ref{#1}}
\newcommand{\secref}[1]{Sec.~\ref{#1}}

\newcommand{\mat}[2]{
\left(\begin{array}{#1}
#2
\end{array}
\right)}
 
\newcommand{\su}[1]{{SU(#1)}}
\newcommand{\GL}[1]{{GL(#1,\mathds{C})}}
\newcommand{\mpol}[2]{\mathrm{M}(#1;#2)}

\newcommand{\hilb}{ \mathcal{H}}
\newcommand{\csp}[1]{\mathbbm{C}^{#1}}

\newcommand{\Rsl}{\hat{\mathcal{M}}}
\newcommand{\Rper}{\hat{\mathcal{S}}}
\newcommand{\RSP}{\widehat{\mathcal{M S}}}
\newcommand{\guni}{{g}}
\newcommand{\gper}{{s}}

\newcommand{\uni}{\hat{\mathcal{U}}}

\newcommand{\mul}{\xi}

\newcommand{\repn}{\psi}
\newcommand{\spin}{spin-\textonehalf\ }
\newcommand{\nspin}{$N$ spin-\textonehalf\ }
\newcommand{\nspinj}{$N$ spin-$J$ }

\title{Geometry of pure states of \nspinj system}
\author{Piotr Kolenderski \\
{\footnotesize\it Institute of Physics, Nicolaus Copernicus University,}\\ {\footnotesize\it Grudzi\k{a}dzka 5, 87-100 Toru{\'n}, Poland \& kolenderski@fizyka.umk.pl}\\}

\begin{document}
\maketitle
\begin{abstract}
    We present the geometry of pure states of an ensemble of \nspinj systems using a generalisation of the Majorana representation. The approach is based on Schur-Weyl duality that allows for simple interpretation of the state transformation under the action of general linear and permutation groups. We show an exemplary application in theory of decoherence free subspaces and noiseless subsystems.
\end{abstract}


\section{Introduction}

The geometrical aspects of physical theories draw attention in fields ranging from classical mechanics through the general relativity to quantum mechanics. The celebrated Bloch sphere picture of a two-level system has its natural application in quantum information theory and quantum optics delivering elegant way of understanding a great number of physical phenomena. Attempts have been made to generalize the Bloch sphere approach to higher dimensional systems \cite{Kimura2005,Chruscinski2009}. The Hopf fibration leading to a nice geometrical structure of one and two qubits has been proposed and applied in the theory of entanglement measures \cite{Levay2005,Levay2006,Chruscinski2006,Brody2007,Paz-Silva2007}. Nevertheless none of the above approaches deliver a simple and general geometrical picture for a ensemble of $N$ spins $J$.

The Majorana representation \cite{Majorana1932,Lee2002,Zimba2006,Bijurkar2007} gives a simple and elegant geometry of quantum states and offers an easy interpretation of the state transformations for spin-$J$ pure states and the  symmetric states of \nspin particles. The representation allowed to gain deeper insight into particular problems of inert states of spinor condensates \cite{Makela2007} and local estimation of Cartesian reference frames \cite{Kolenderski2008b}. The basic idea of the Majorana representation is that the spin-j state can be uniquely represented as $2j$ points on the unit sphere. The positions of the points on the sphere are easy to compute as roots of a certain polynomial. The beauty of the approach expresses in the fact that under the action of $\su{2}$ matrix all the points rotate as a solid body. 

A modification of the Majorana approach for $N$ qubits has been applied in context of separability problem \cite{Makela2009} and allowed to find the geometry of separable states.  The method is based on observation that a state of \nspin can be regarded as a state of $2^N$ level system. Nevertheless, this approach lacks the desired behavior under the special unitary matrix action as the respective points does not transform in the simple way. 

Here we present the geometry of pure states of an ensemble of \nspinj system, which is a direct generalization of the celebrated Bloch sphere for a single qubit and has analogical characteristics with respect to the unitary matrix transformations. The action of the unitary and permutation group is also discussed. Furthermore, we show an exemplary application of the method in the context of decoherence free subspaces (DFS) and noiseless subspaces (NS) for $N$ qubit system (\nspin). The presented geometry allows to distinguish between the logical and physical states of the system and yields further  insight into the nature of quantum operations in DFS/NS.

The paper is organized as follows. In \secref{sec:majorana} we recall the Majorana representation and present its exemplary application in quantum phase estimation, entanglement classification under stochastic local operations and classical communication (SLOCC) and quantum optics. Next, in \secref{sec:NSpinJ}, we introduce the geometry of \nspinj states based on the Schur-Weyl duality and discuss the general linear and permutation group transformation. The exemplary application of the method for DFS/NS theory is also given.

\section{The geometry of spin-$J$ states}\label{sec:majorana}
\subsection{Majorana representation}
First, let us briefly recall the Majorana representation \cite{Majorana1932, Prenrose1984,Coecke1998,Bijurkar2007}, which  allows one to uniquely represent spin-$J$ state as $2J$ points on the unit sphere. The method is a direct generalization of the Bloch sphere for \spin particle. For an arbitrary state $\ket{z}=[\cos(\theta/2),\sin(\theta/2)\exp(i \phi)]$ a stereographic projection can be used instead of a Bloch vector $\mathbf{n}=[\sin\theta \cos\phi, \sin\theta\sin\phi,\cos\theta]$. This way the state of a \spin can be parameterized with a single complex number $z=e^{-i \phi}\cot{\theta/2}$, where $z= \infty$ for $\theta=0$. Next, let $\ket{z_{\perp}}$ be a state orthogonal to $\ket{z}$, then for a given state of spin-$J$:
\begin{equation}
    \ket{\psi}=\sum_{m=-J}^{J} \psi_m \ket{J,m}
\end{equation}
an overlap $\bra{z_\perp}^{\otimes 2J}\ket{\psi}$ 
is proportional to the {\em Majorana polynomial}:
\begin{equation}\label{eq:MajoranaPolymnomial}
    \mpol{\ket{\psi}}{z}=\sum_{m=-J}^{J}(-1)^{k} \mat{c}{2J \\ J + m}^{\frac{1}{2}} \psi_m  z^{J+m} 
\end{equation}
up to an irrelevant function of $z$ having no roots. Then by the fundamental theorem of algebra, the polynomial $\mpol{\ket{\psi}}{z}$ can be uniquely factored. In consequence for each spin-$J$ state there exist a unique set of $2J$ complex numbers composed of $\tilde{N}$ roots of the Majorana polynomial $\{z_1,z_2,\dots,z_{\tilde N} \}$ supplemented by ($2J-\tilde N$)-element set of $\infty$. Each element of the set corresponds to a \spin state, thus there is one to one correspondence between the spin-$J$ state and $2J$ \spin states, which can be represented as points on a Bloch sphere. In principle if a certain state occurs $d$ times we shall refer to such a state and a corresponding point as to  \emph{d-fold degenerate}.
Moreover, note that the method can be applied also to the totally symmetric states of \nspin\  as they are related to the spin-$N/2$ states.

The Majorana representation, however, cannot be simply generalized for mixed states. For \spin states the points inside the Bloch ball correspond to all possible mixed states. This idea cannot be easily transferred for spin-$J$ states with $J>1/2$, which is the direct conclusion of the state parameter counting: In general a mixed state of spin-$J$ is parameterized by $(2J+1)^2-1$ real parameters whereas $2J$ points inside the ball are fully described by $6J$ real numbers. One can see that the equality is only for $J=1/2$ and in general the number of the mixed state parameters is much grater than the number of parameters for $2J$ points in the Bloch ball.

\subsection{State transformation}
The geometry associated with the Majorana representation has beautiful properties with respect to the transformations of the invertible matrices with a nonzero determinant that comprise the general linear group  $\GL{2}$. It is straightforward to see that, when the matrix representation of $\GL{2}$ acts on the spin-$J$ state, all $2J$ \spin states undergo the same transformation.
Indeed, solving the relevant Majorana polynomial $\mpol{\ket{\psi}}{z}$ is equivalent to searching all the \spin states $\ket{z_\perp}^{\otimes 2J}$ that are perpendicular to the state $\ket{\psi}$. Hence, when the spin-$J$ state is transformed $\Rsl_J\ket{\psi}$ it is equivalent with a transformation of the \spin states $\Rsl_{1/2}\ket{z}$ as $\bra{z_\perp}^{\otimes 2J}\Rsl_J \ket{\psi}=(\bra{z_\perp}\Rsl_{1/2}^\dagger)^{\otimes 2J}\ket{\psi}$. Here $\Rsl_J$ denotes irreducible $2J+1$ dimensional representation of $\GL{2}$.

As all $2J$ \spin states undergo the same $\GL{2}$ action let us look at the transformation closer. To do so recall that any matrix $\Rsl \in \GL{2}$ can be uniquely decomposed as $\Rsl = \uni \hat{\mathcal{R}}$, where $\uni \in U(2)$ is a unitary matrix and $\hat{\mathcal{R}}$ is a hermitian positive semidefinite matrix. The action of unitary matrix is trivial as it is a simple rotation of Bloch sphere, hence the unitary transformation of spin-$J$ state is reflected in the rotation of the corresponding points all together as a rigid solid. The hermitian matrix transformation requires more attention. Note that the resulting state is not normalized hence its Bloch vector neither. The hermitian matrix transforms the sphere into an ellipsoid and moves it in the certain direction in such a way that the center of the sphere is always inside the resulting ellipsoid. Next, the normalization procedure amounts to shrink or lengthen the Bloch vectors. In result, starting form unit sphere with the uniform state density (according to the Haar measure) the hermitian positive semidefinite matrix transforms it to the unit sphere with the modified state density. It thickens the states in the neighborhoods of two antipodal points in the  direction characterized by the hermitian matrix eigenvectors. 

Hence, the general linear group $\GL{2}$ when acting on the state of spin-$J$ changes the relative orientation of its points. Nevertheless, it is not possible to transform arbitrary state into any other this way as $\GL{2}$ does not has enough degrees of freedom. Any points combination on the Bloch sphere can be transformed into any other only using $\su{2J+1}$ group.

\subsection{Applications}
The Majorana representation turned out to be very useful in a great number of problems. The inert states of spinor condensates \cite{Makela2007} and the optimal states for local reference frame estimation \cite{Kolenderski2008b} has been found to be related to  Platonic solids. We recall below other problems that have a simple interpretation in terms of Majorana representation -- phase estimation, SLOCC entanglement classes characterization \cite{Bastin2009} and the multi-photon states generation in a process of spontaneous parametric down conversion (SPDC) \cite{McCusker2009}.

\subsubsection{Phase estimation}
In quantum estimation theory one considers the state that depends on the set of parameters $\hat{T}(p_1,p_2,\dots,p_k)\ket{\psi}$, where $\hat T$ is a given transformation and $\ket{\psi}$ is a state of the system. The question is: what is the optimal state $\ket{\psi}$ that allows for the best estimation of small deviation of parameters from their given initial values $p_{1}^{(0)},p_{k}^{(0)},\dots,p_{k}^{(0)}$, see Ref.~\cite{Helstrom1976}.

\begin{figure}[ht!]
  \begin{tabular}{ccc}
\subfigure[$\ket{\psi_1}$] {\includegraphics[width=0.3\columnwidth]{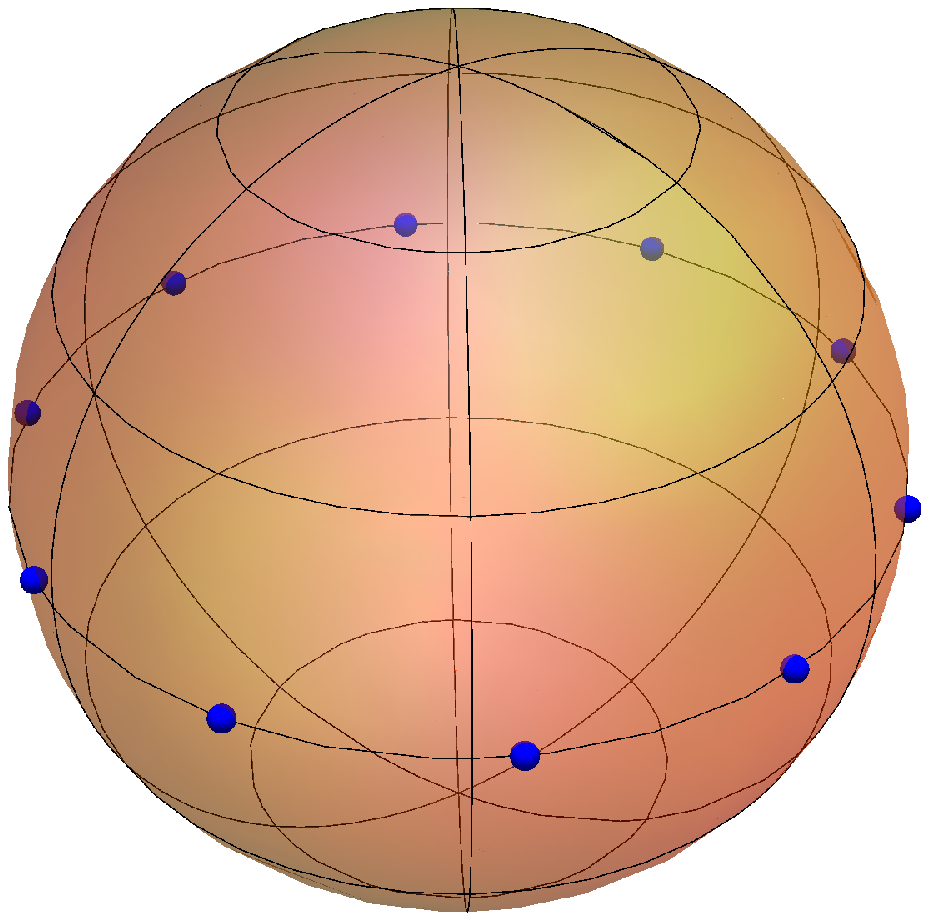}\label{fig:state:NOON}} &\subfigure[$\ket{\psi_2}$] {\includegraphics[width=0.3\columnwidth]{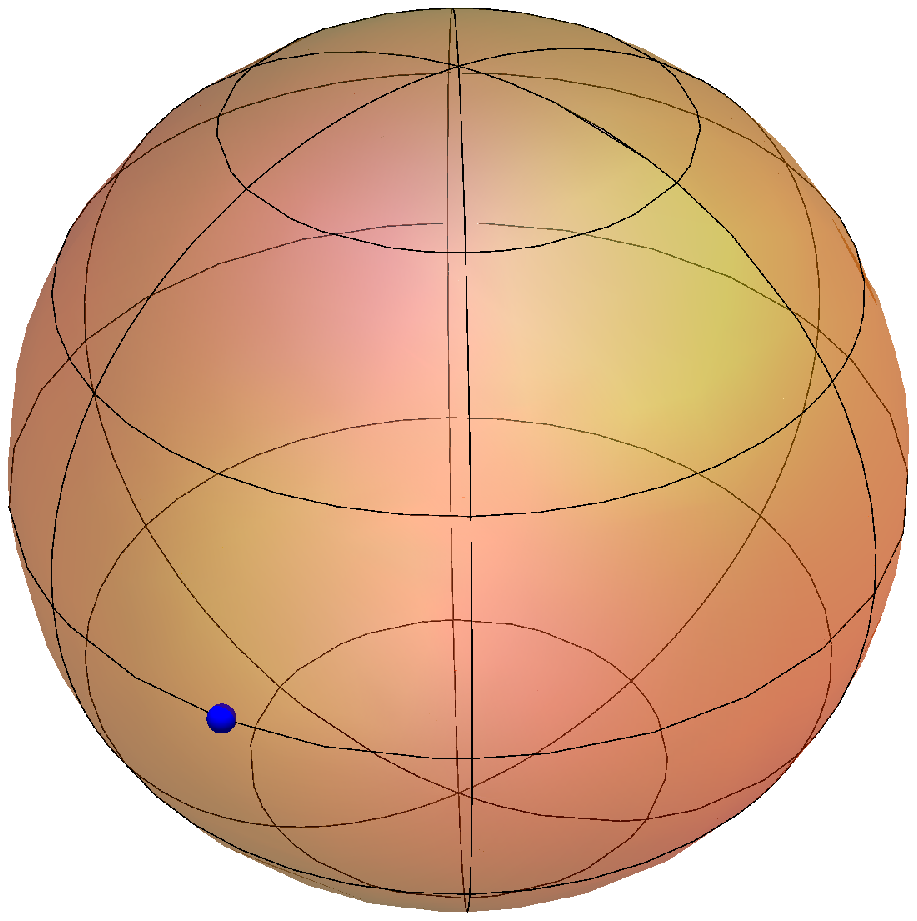}\label{fig:state:ShotNoise}}
&\subfigure[$\ket{\psi_3}$] {\includegraphics[width=0.3\columnwidth]{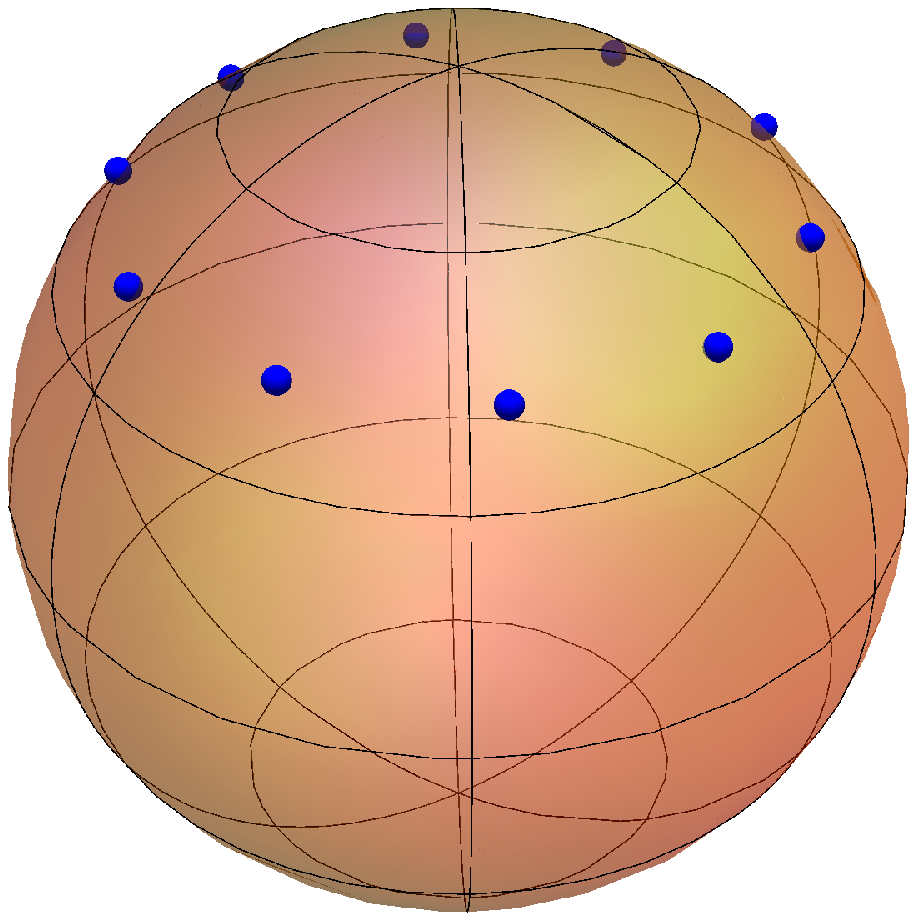}\label{fig:state:aNOON}}
  \end{tabular}
  \caption{Representation spheres for: \subref{fig:state:NOON} the NOON state $\ket{\psi_1}$, \subref{fig:state:ShotNoise} the state $\ket{\psi_2}$ leading to shot noise limit  and \subref{fig:state:aNOON} the modification of NOON state $\ket{\psi_3}$  for $N=12$.}
\label{fig:states}
\end{figure}

 In particular the question can be put as follows: What is the optimal $N$ qubit state for a phase estimation? In other words the state $\ket{\psi}$ must be found such that $\exp(i \phi \hat{\sigma}_z )^{\otimes N}\ket{\psi}$ is the most sensitive for the small changes of the phase $\phi$ from its initial value $\phi^{(0)}=0$.  The standard notation for Pauli matrix has been used $\hat{\sigma}_z$. The answer for the question is the NOON \cite{Bollinger1996,Dowling1998}  state: 
\begin{equation}
	\ket{\psi_1}=\left(\ket{0}^{\otimes N}+\ket{1}^{\otimes N}\right)/\sqrt{2}
\end{equation}
that leads to the Heisenberg limit \cite{Yurke1986}. It can be rewritten in the spin notation  as a superposition of spin-$N/2$ up and spin-$N/2$ down: $\ket{\psi_1}=\left(\ket{N/2,N/2}+\ket{N/2,-N/2}\right)/\sqrt{2}$. It is easy to see that the NOON state $\ket{\psi_1}$ corresponds to $N$ equally spaced points on the equator, see \figref{fig:states}(a). 

The Majorana representation allows one to gain further insight into the nature of the optimality of the NOON state. The unitary transformation  $\exp(i \phi \hat{\sigma}_z )^{\otimes N}$ corresponds to the  rotation of all the points with respect to the $z$ axis by an angle $\phi$. When searching for the optimal state one should prefer those which change the most with respect to small phase change. Let us compare here the geometries of the NOON state with the state leading to the shot noise limit \cite{Caves1981}:
\begin{equation}
	\ket{\psi_2}=\frac{1}{\sqrt{2^N}}(\ket{0}+\ket{1})^{\otimes N},
\end{equation}
which is represented as a single $N$-fold degenerated point on equator, see \figref{fig:states}(b). The NOON state is transformed to itself after $\phi=2\pi/N$, whereas $\ket{\psi_2}$ after the full $\phi=2\pi$ rotation, thus $\ket{\psi_1}$ is more sensitive for the rotations than $\ket{\psi_2}$. Hence comparing NOON state $\ket{\psi_1}$ with the state leading to shot noise limit $\ket{\psi_2}$, the former one fails the competition. Moreover, the geometry allows to easily see that using the NOON state the phase can be estimated in the range $(0,\pi/N)$ as the points are transformed into themselves for every $\pi/N$ rotation.

In order to gain more intuitions, we consider the modification of the NOON state:
\begin{equation}
	\ket{\psi_3}=\alpha\ket{0}^{\otimes N}+ \beta\ket{1}^{\otimes N}.
\end{equation}
which is represented by the $N$ equally spaced points on the circle in the plain parallel to equator, what is depicted in see \figref{fig:states}\subref{fig:state:aNOON}. The position of the circle depends on the coefficients $\alpha$, $\beta$ and the number of particles $N$.
One can see that, when $\alpha=\beta=1/\sqrt{2}$, then it is a NOON state: $\ket{\psi_1}$ and when $\alpha=0$, then it is a state of total angular momentum $N/2$ and projection $N/2$: $\ket{N/2,N/2}$. In Majorana representation the former situation corresponds to one $N$ fold degenerate point placed on the North pole. Now it is easy to see, that the state $\ket{\psi_3}$ is less sensitive for the unitary rotation $\exp(i \phi \hat{\sigma}_z )^{\otimes N}$ than NOON state as in the limit of $\alpha=0$ it converges to the  $\ket{N/2,N/2}$, which is immune for the rotation around $z$ axis.


\subsubsection{SLOCC entangled classes }
Recently Bastin et al.~\cite{Bastin2009} have solved the problem of the entanglement classification under SLOCC  for symmetric $N$ qubit states. The problem is to find the classes of symmetric states that are connected via invertible local transformations. In other words, two states $\ket{\psi}$ and $\ket{\phi}$ are said to belong to the same class if and only if there exist $\Rsl \in \GL{2}$ such that $\ket{\psi}= \Rsl ^{\otimes N} \ket{\phi}$. 

The problem and its solution can be simply understood within Majorana representation. Two states $\ket{\psi}$ and $\ket{\phi}$ refer to two configurations of the points on the Bloch sphere. Using the invertible local transformations one can transform the corresponding points. However, only for states associated with 3 points it is possible co transform any configuration into any other as the $\GL{2}$ group is parameterized with 7 real numbers. On the other hand  by $\GL{2}$ group one cannot change the degeneracy of the points as it acts in the same way on each point, hence it cannot split any degenerate one. Hence the necessary condition for the states to be in the same SLOCC entanglement class in the language of Majorana representation is the same number identically degenerated points.

\subsubsection{Quantum optics}

McCusker and Kwiat in Ref.~\cite{McCusker2009} have proposed a method of producing multi-photon states. The scheme is based on repeated SPDC process where one photon of the pair heralds the presence of the other which is then stored in optical cavity. By repeating the process of adding the photons and manipulating their polarizations, the state $\ket{\psi}$, which is a product of an arbitrary polarizations, can be built up:
\begin{equation}\label{eq:QuantumOptics}
	\ket{\psi} = \prod_{n=0}^{N-1} (\alpha_n \hat{a}_H^\dagger +\beta_n \hat{a}_V^\dagger)\ket{\text{vac}}
\end{equation}

The multi-photon state, which is a superposition of two polarizations in a single spatiotemporal mode is in one-to-one correspondence with a spin state. This relation is know as the Schwinger representation. The mapping can be easily done by simple change of representation switching from the states $\ket{n_H,n_V}$ of definite number $n_H$ ($n_V$) of horizontally (vertically) polarised photons to the states $\ket{(n_H+n_V)/2,(n_H-n_V)/2}$ of definite sum and difference of polarization occupation numbers. The sum divided by two corresponds to a total angular momentum and the difference divided by two to its projection.

The $N$ photon state \eqref{eq:QuantumOptics} corresponds to a spin-$N/2$, hence it can be represented on the Bloch sphere via Majorana representation. Moreover, the orientation of the points are given by the Bloch vectors of the states $(\alpha_n \hat{a}_H^\dagger +\beta_n \hat{a}_V^\dagger)\ket{\text{vac}}$. In consequence the experimental process of state construction by consecutive single photon addition is reflected in the process of addition of new points on the Bloch sphere.

\section{Geometry of \nspinj states} \label{sec:NSpinJ}
The Majorana representation can be used for N spin-$J$ systems, when the state is regarded as a $(2J+1)^N$ level system and as such can be represented as $(2J+1)^N-1$ points on the Bloch sphere \cite{Bijurkar2007}. This approach allowed to write corresponding Majorana polynomial \cite{Makela2009} and formulate the separability criteria associated with an elegant geometry of separable states. However the action of unitary matrix leads to the highly nontrivial behavior of the points on the Bloch sphere. We discuss here the \nspinj state geometry that overcomes this problem and allows for simple interpretation of state transformation under the general linear $\GL{2J+1}$ and permutation $S_N$ groups action.

\subsection{The representation}
The approach is based on the Schur-Weyl duality. We consider here the permutation $S_N$ and general linear $\GL{2J+1}$ group and its representations $\Rper$ and $\Rsl$ over the Hilbert space of \nspinj states. The representation of permutation group  for a given element $s \in S_N$ is given by:
\begin{equation}
	\Rper(\gper) \ket{a_1}\ket{a_2}\dots\ket{a_N}= \ket{a_{s(1)}}\ket{a_{s(2)}}\dots\ket{a_{s(N)}}
\end{equation}
It refers to an interchange of the respective single particle states. Moreover, the action of the representation of the general linear group element  $\guni \in \GL{2J+1}$ is an action of the group on each particle:
\begin{equation}
	\Rsl(\guni) \ket{a_1,a_2,\dots,a_N}= \Rsl(\guni)\ket{a_1} \Rsl(\guni)\ket{a_2} \dots \Rsl(\guni)\ket{a_N}
\end{equation}
Next, as the representations commute we consider the joint action of both representations, which we will denote by $ \RSP(\guni,\gper) =\Rsl(\guni) \Rper(s)$. The Schur-Weyl theorem states that the representation of joint action of the general linear and the permutation groups  $\GL{2J+1}\times S_N$ can be decomposed into irreducible representations in the following way:
\begin{equation}\label{eq:ShurWeyl}
   \RSP(\guni,\gper) \cong \bigoplus_{\lambda \in \text{Par}(N,d)}\Rsl_\lambda(\guni)\otimes \Rper_\lambda(\gper),
\end{equation}
where  $\Rsl_\lambda(\guni)$ and $\Rper_\lambda(\gper)$ are irreducible representations (irreps) of $\GL{2J+1}$ and $ S_N$, respectively, and Par$(N,d)$ is a set of all partitions of $N$ into $d$ parts. In conjunction with the decomposition of \eqref{eq:ShurWeyl} the following decomposition of the Hilbert space of the system of N spins-$J$ can be done:
\begin{equation}\label{eq:gen:space}
    \hilb_J^{\otimes N} = \bigoplus_{\lambda \in \text{Par}(N,d)} \hilb_\lambda^{GL} \otimes \hilb_{\lambda}^{S}
\end{equation}
where $\hilb_\lambda^{GL}$ and $\hilb_{\lambda}^{S}$ are spaces where the irreps $\Rsl_\lambda$ and $\Rper_\lambda$ act, respectively. The dimensions of the subspaces $\hilb_\lambda^{GL}$ and $\hilb_{\lambda}^{S}$ can be computed using Young diagrams method. Next, with the Hilbert space decomposition in hand one can easily express an arbitrary state of \nspinj system $\ket{\Psi} \in \hilb_J^{\otimes N}$ as:
\begin{equation}\label{eq:gen:state}
    \ket{\Psi} = \sum_{\lambda,\alpha} \xi_\lambda^\alpha \ket{\psi_\lambda^\alpha}_\lambda\otimes\ket{\alpha}_\lambda,
\end{equation}
where $\ket{\psi_\lambda^\alpha}_\lambda\in \hilb_\lambda^{GL}$ and $\ket{\alpha}_\lambda\in\hilb_{\lambda}^{S}$. We assume that the states $\ket{\psi_\lambda^\alpha}_\lambda$ are normalized to unity in such way that all have the same global phase and $\ket{\alpha}_\lambda$ are orthonormal $_\lambda\langle \alpha \ket{\beta}_\lambda=\delta_{\alpha\beta}$. The above decomposition is unique therefore there is one-to-one correspondence between the state of \nspinj and the set  $\left\{\ket{\psi_\lambda^\alpha}_\lambda\right\}_{\lambda,\alpha} \cup \left\{\ket{\xi}\right\}$, where $\ket{\xi}=\bigoplus_\lambda \sum_\alpha \xi_\lambda^\alpha \ket{\alpha}_\lambda$. 
We will refer to states form the set $\left\{\ket{\psi_\lambda^\alpha}_\lambda\right\}_{\lambda,\alpha}$ as \emph{representation states} and analogically refer to $\ket{\xi}$ as \emph{multiplicity state}. Furthermore, if for some $\alpha$ and given $\lambda$ representation states are identical we will refer to the corresponding state as \emph{degenerate} one.

Next, each of the states can be easily represented on the Bloch sphere via Majorana representation. The states $\left\{\ket{\psi_\lambda^\alpha}_\lambda\right\}_{\lambda,\alpha}$ can all be drawn on \emph{representation sphere} using different colors to distinguish between different $\alpha$ and $\lambda$. The state $\ket{\mul}$ can be drawn on the separate \emph{ multiplicity sphere}. Note that, the split is done due to fundamental difference between the representation and multiplicity states in context of the action of the group $\GL{2J+1}\times S_N$. 

\subsection{Properties}
Let us look now how the state transformation is reflected in its geometry on the representation and multiplicity  spheres. First, we consider a separable  state of a given $\lambda$ and $\alpha$: $\ket{\psi}_\lambda\otimes\ket{\alpha}_\lambda$, where $\ket{\psi}_\lambda \in \hilb^{GL}_\lambda$ and $\ket{\alpha}_\lambda \in \hilb^{S}_\lambda$. The geometry of such a state is particularly simple as the multiplicity state $\ket{\xi}=\ket{\alpha}_\lambda $ and therefore there is only one representation state $\ket{\psi}_\lambda$. The action of the representation $ \RSP$ on the state is given by:
\begin{equation}
     \RSP(\guni,\gper) \ket{\psi}_\lambda\otimes\ket{\alpha}_\lambda = \Rsl_\lambda(\guni) \ket{\psi}_\lambda \otimes \Rper_\lambda(\gper)\ket{\alpha}_\lambda
\end{equation}
The special linear group representation acts only on the representation state whereas the permutation group representation on the multiplicity state. However, the permutation group in general modifies the degeneracy of the representation state as after the action of $ \RSP(\guni,\gper)$ the multiplicity state is given by $\RSP(\guni,\gper)\ket{\xi}=\sum_{\alpha'}s^{\alpha \alpha'}_\lambda\ket{\alpha'}_\lambda$. The degeneracy of $\ket{\psi}_\lambda$ is equal to a number of nonzero amplitudes $s^{\alpha \alpha'}_\lambda$.

It is easy to see that the degeneracy change caused by the permutation group action in consequence leads to the modification of the representation states. Indeed, when acting with the representation of identity element for general linear group and arbitrary element $\gper$ of permutation group on general \nspinj state $\ket{\Psi}$:
\begin{equation}
     \RSP(1,\gper)\ket{\Psi} = \sum_{\lambda,\alpha} \left(\sum_{\alpha'}\xi_\lambda^{\alpha'}s_\lambda^{\alpha' \alpha} \ket{\psi_\lambda^{\alpha'}}\right) \otimes \ket{\alpha}_\lambda,
\end{equation}
it is seen that the new representation states are proportional to $\ket{\tilde \psi_\lambda^\alpha}_\lambda \propto \sum_{\alpha'}\xi_\lambda^{\alpha'}s_\lambda^{\alpha' \alpha} \ket{\psi_\lambda^{\alpha'}}$. In general $S_N$ mixes coherently the representation states.

On the other hand, the representation of an arbitrary element of general linear group and identity element of permutation group $\RSP(\guni,1)$ on the state:
\begin{equation}
     \RSP(\guni,1)\ket{\Psi} = \sum_{\lambda,\alpha} \xi_\lambda^{\alpha} \left(\Rsl_\lambda(\guni) \ket{\psi_\lambda^{\alpha}}\right) \otimes \ket{\alpha}_\lambda,
\end{equation}
resort to the transformation of only representation states.

\subsection{Applications}
The presented method is particularly useful in the theory of decoherence free subspaces and noiseless subsystems. The problem is following. Assume we have $N$ qubits, which experience an unknown unitary rotation $\uni^{\otimes N}$, where $\uni \in \su{2}$. The question is how to encode a logical state into physical qubits such that the logical state does not change after arbitrary unitary rotation  $\uni^{\otimes N}$. The detailed analysis can be found in Refs.~\cite{Kempe2001, Byrd2006, Bishop2009} and here we will discuss the geometric aspects of the problem.

We consider the subgroup $\su{2}$ of general linear group $\GL{2}$. The properties of $\GL{2}$ discussed in the previous section are valid with respect to $\su{2}$. For this special case the decomposition of Hilbert space  according to \eqref{eq:gen:space} amounts to the direct sum of tensor product of a total angular momentum subspace $\hilb_j$ and a multiplicity subspace $\csp{d_j}$. The dimensions of the subspaces are respectively $2j+1$ and $d_j={(2 j+1) \binom{N}{\frac{N}{2}-j}}/{(j+\frac{N}{2}+1)}$. 
Then, in general, the action of  $\uni^{\otimes N}$ on the state of $N$ qubits is an unitary rotation of its representation states: 
\begin{equation}\label{eq:dfs:state}
    \uni(\guni)^{\otimes N}\ket{\Psi} = \sum_{j=(N \bmod 2)/2}^{N/2} \mul_j^\alpha \uni_j(\guni)\ket{\psi_j^\alpha}_j\otimes\ket{\alpha}_j,
\end{equation}
where $\uni_j(g)$ is $2j+1$ dimensional irrep of $\guni\in\su{2}$. In Majorana representation this can be seen as the rotation of all the representation points as a solid body, whereas the points on the multiplicity sphere do not experience any modification. In consequence all the information about the logical state must be encoded in multiplicity sphere. 
Hence all interesting logical qubit dynamics can be investigated there.

As an example, let us look at the simplest DFS for three qubits. The most general form of the state is given by:
\begin{equation}\label{eq:dfs:3}
    \ket{\Psi}=\mul_{3/2} \ket{\repn_{3/2}}+ \mul_{1/2}^0 \ket{\repn_{1/2}^0}+ \mul_{1/2}^1 \ket{\repn_{1/2}^1}
\end{equation}
When the logical qubit is encoded in the spaces of total angular momentum $j=1/2$ it is immune to the strong collective noise. Typically arbitrary states of \spin: $\ket{\psi_{1/2}^0}$ and $\ket{\psi_{1/2}^1}$ can represent logical $0$ and $1$. The logical qubit can be encoded entirely in the multiplicity state:
\begin{equation}\label{eq:dfs:3:logical}
    \ket{\Psi_L}=\mul_{1/2}^0 \ket{ \repn_{1/2}^0}+ \mul_{1/2}^1 \ket{\repn_{1/2}^1}.
\end{equation}
Then its multiplicity state is given by $\ket{\mul_L}=(0,\mul_{1/2}^0,\mul_{1/2}^1)$.

As was observed in Ref.~\cite{Bishop2009} for 3 \spin system, the hamiltonians that can be used for physical realisation of unitary transformation of logical qubit can be constructed based on  the algebra of quantum operators $\hat{X}_L,\hat{Y}_L,\hat{Z}_L$. The operators are  a linear combinations of physical spins permutations. For example the logical $\hat{ Z}_L$ operator is a combination of three permutations:
\begin{equation}
    \hat{ Z}_L =\frac{1}{3}(\Rper_{3214}+\Rper_{1324}-2\Rper_{2134})
\end{equation}
where $\Rper_{i_1 i_2 i_3 i_4}$ denotes the permutation operator which changes physical qubit number 1 with $i_1$, number 2 with $i_2$ and so on. In consequence the SU(2) rotation of logical qubit can be obtained by\cite{Bishop2009}:
\begin{equation}
	\uni_L=\exp\left(i \alpha \hat{Z}_L\right) \exp\left(i \beta \hat{Y}_L\right) \exp\left(i \gamma \hat{Z}_L\right)
\end{equation}
where $\alpha$, $\beta$ and $\gamma$ are the Euler angels. 

\begin{figure}[h]
	
\centering
\frame
{
\subfigure[$\ket{\Psi_L}$] {
\begin{tabular}{lr}
\includegraphics[width=0.2\columnwidth]{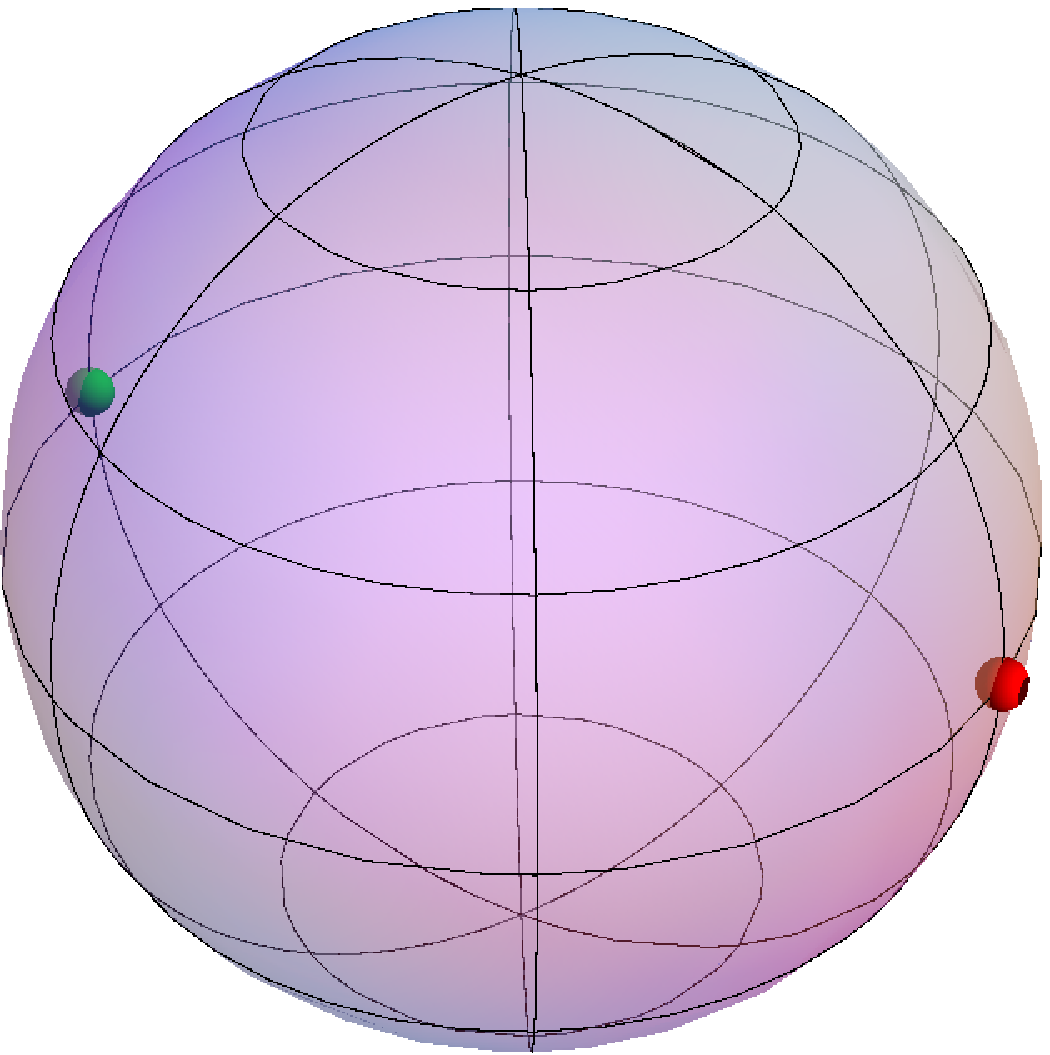} &
\includegraphics[width=0.2\columnwidth]{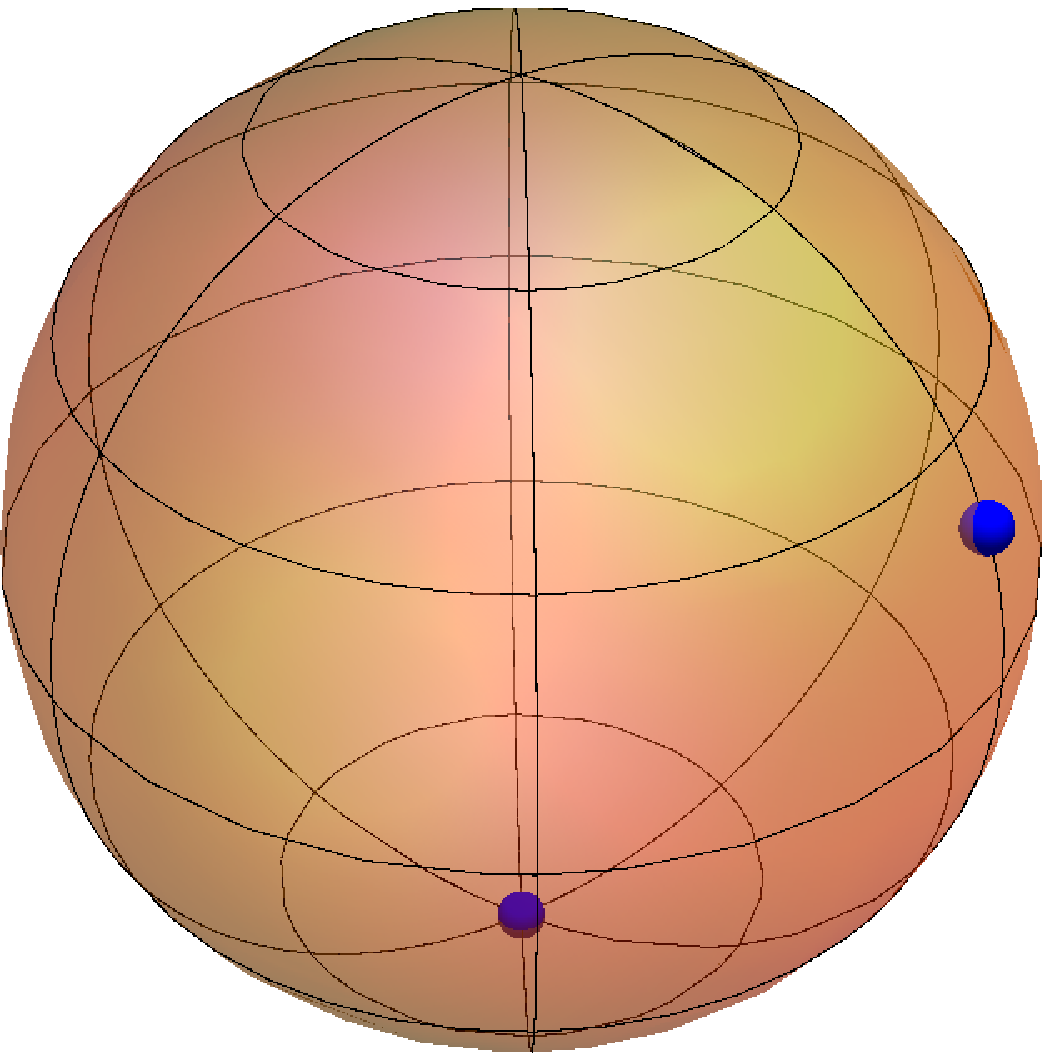}
\end{tabular}\label{fig:dfs}
}}\\
\hspace{1cm}\\
\frame{
\subfigure[$\uni^{\otimes 3}\ket{\Psi_L}$] {
\begin{tabular}{lr}
\includegraphics[width=0.2\columnwidth]{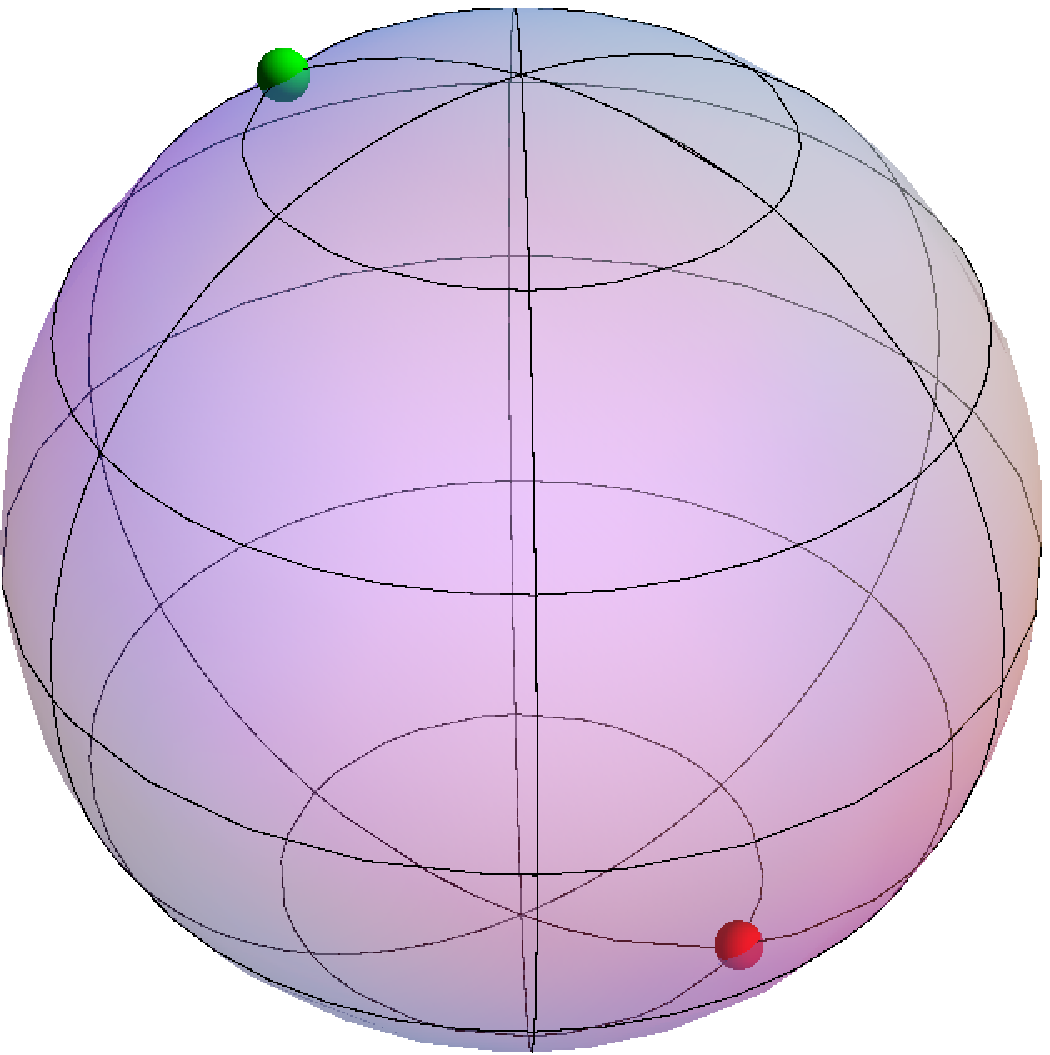} &
\includegraphics[width=0.2\columnwidth]{diagm1}
\end{tabular}\label{fig:dfs:noise}
}}
\frame{
\subfigure[$\uni_L\ket{\Psi_L}$] {
\begin{tabular}{lr}
\includegraphics[width=0.2\columnwidth]{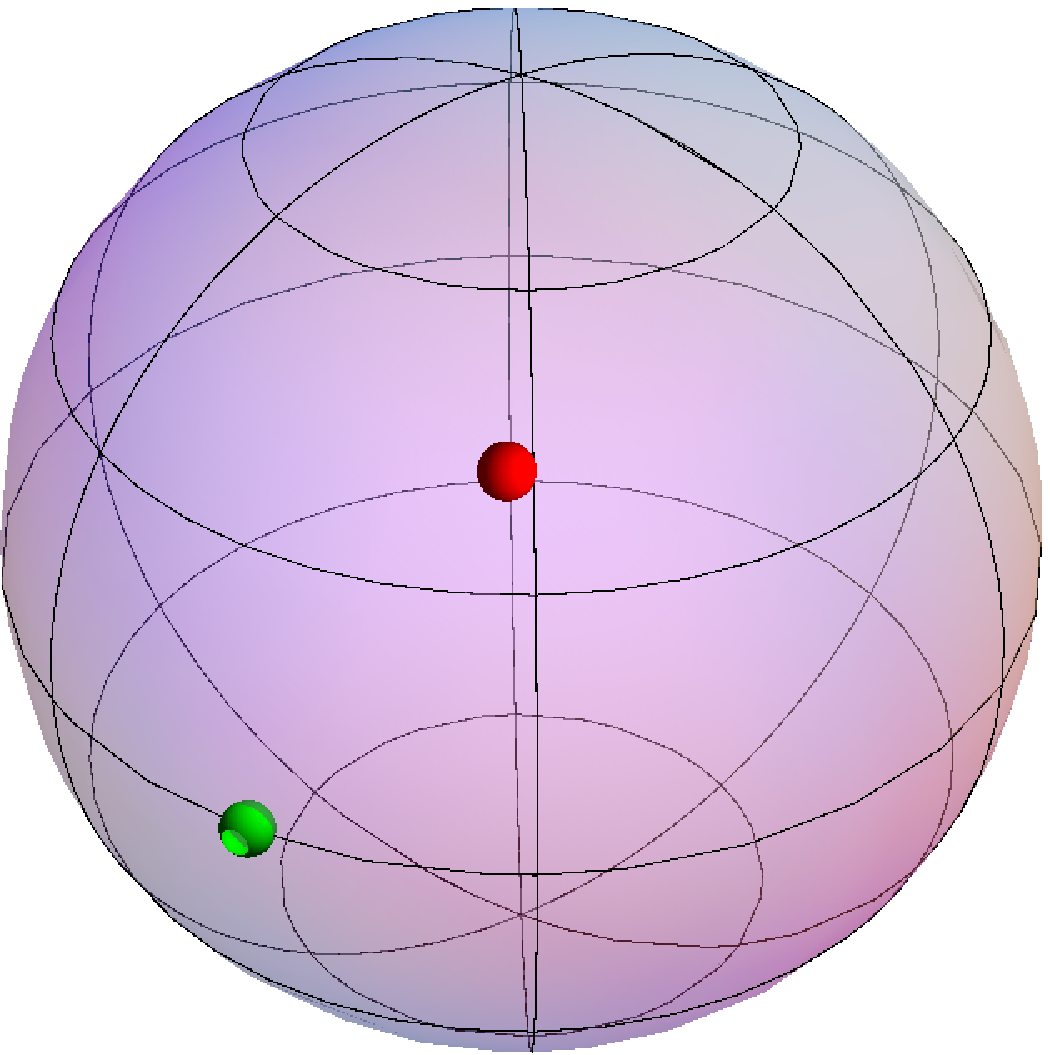} &
\includegraphics[width=0.2\columnwidth]{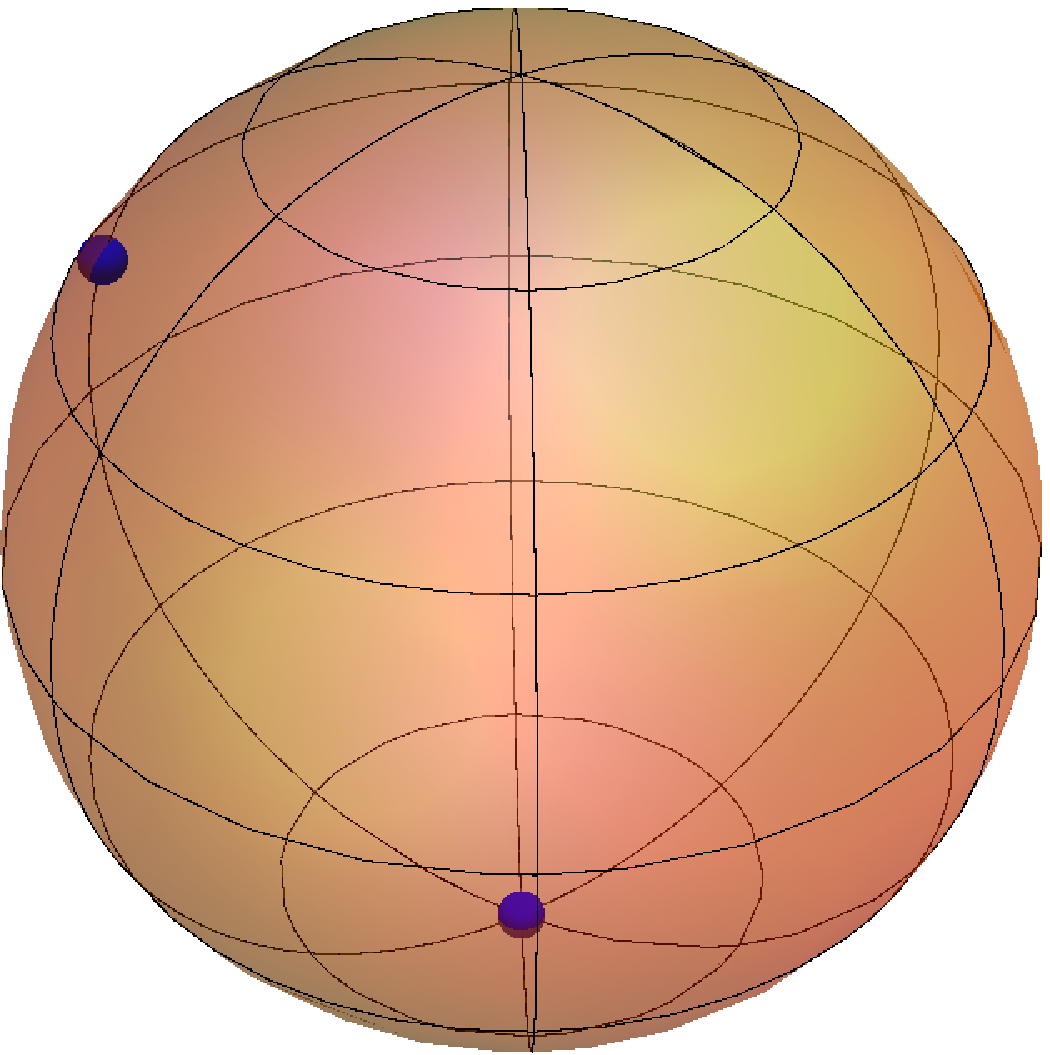}
\end{tabular}\label{fig:dfs:log}
}}
\caption{a) The geometry of an exemplary state $\ket{\Psi_L}$ depicted in the representation (left) and multiplicity (right) spheres. b) Under the action of $\uni^{\otimes 3}$ only the representation sphere experiences a modification. c) The logical qubit transformation, in general, changes the representation states.}
\label{fig:diagram}
\end{figure}

A diagram summarising the presented discussion is depicted in \figref{fig:diagram}. For an exemplary state:
\begin{eqnarray}\label{eq:dfs:example} \ket{\Psi_L}=\frac{1}{2\sqrt{6}}\left(2(\ket{110}+\ket{001})-(1+\sqrt{3})(\ket{101}+\ket{100})+\right. \\ \left.
    \nonumber (-1+\sqrt{3})(\ket{011}+\ket{010})\right)
\end{eqnarray}
one can easily find the multiplicity state: 
\begin{equation}
	\ket{\xi}=\left(0,\frac{1}{\sqrt{2}},\frac{1}{\sqrt{2}}\right)
\end{equation}
and representation states: 
\begin{eqnarray}
	\ket{\psi_{1/2}^0}_{1/2}=\frac{1}{\sqrt{2}}\left(\ket{0}+\ket{1}\right)\\
	\ket{\psi_{1/2}^1}_{1/2}=\frac{1}{\sqrt{2}}(\ket{0}-\ket{1}).
\end{eqnarray}

The state $\ket{\Psi_L}$ is depicted in \figref{fig:diagram}\subref{fig:dfs}: the representation states are presented on the left sphere and the multiplicity state on the right one. Under the action of arbitrary $\uni^{\otimes 3}$ only the representation sphere experiences modification. The logical qubit is immune for this kind of operation as it can be seen in the lower left box in \figref{fig:diagram}. Moreover, let us consider a simple unitary rotation of the logical qubit around the $z$ axis $\uni_L=\exp\left(i \alpha \hat{ Z}_L\right)$. It is easy to check that this transformation modifies the multiplicity state: $\uni_L (0,\mul_{1/2}^0,\mul_{1/2}^1)=(0,\mul_{1/2}^0 e^{i \alpha},\mul_{1/2}^1 e^{-i \alpha})$, what is depicted in the right box in \figref{fig:diagram}  for $\alpha=\pi$. The logical qubit transformation $\uni_L$ in general changes the orientation of the points on both spheres.

\section{Conclusions}
We discussed the Majorana representation, which allows one to represent arbitrary pure state of multilevel system as points on Bloch sphere, which are rotated as a rigid body under the action of special unitary group \su{2}. The method cannot be considered as a tool providing the solution. However, it proved to be very useful offering deeper insight and understanding of the problem. 

The main result presented in Sec.~3 was a generalisation of the Majorana representation for the pure states of \nspinj systems. When applied to the theory of decoherence free subspaces, it allowed as to geometrically separate the noisy dynamics and the logical state transformation.

The main drawback of the Majorana representation and presented geometry of the states \nspinj systems is that both work only for pure states. Hence, in is very desirable to construct the mixed states geometry that allows one for simple understanding of the problem under consideration.


\section*{Acknowledgements}
I acknowledge insightful conversations with Konrad Banaszek and Rafa{\l}  Demkowicz-Dobrza\'{n}ski. This work has been supported by Polish MNiSW (N N202 1489 33) and the Future and Emerging Technologies (FET) programme within the
Seventh Framework Programme for Research of the European Commission, under the FET-Open grant agreement CORNER no. FP7-ICT-213681 and the Operational Programme Human Capital under the project Krok w przysz{\l}o\'{s}\'{c} II and ZPORR project.

\bibliographystyle{abbrv}

\end{document}